%% file: main.tex
\documentclass[a4paper]{jpconf}
\usepackage{pgf}
\usepackage{graphicx}	% Include figure files
\usepackage{dcolumn}	% Align table columns on decimal point
\usepackage{bm}				% bold math
\usepackage{units}
\usepackage[utf8x]{inputenc}
\usepackage[T1]{fontenc}
\usepackage{subfigure}
\begin{document}

\title{Scanning electron microscopy of Rydberg-excited Bose-Einstein condensates} 
\author{T. Manthey$^1$,T.M. Weber$^1$,T. Niederprüm$^1$,P. Langer$^1$,V. Guarrera$^{1,2}$,G. Barontini$^{1,3}$,H. Ott$^1$}
\address{$^1$Research Center OPTIMAS, Technische Universität Kaiserslautern, Gottlieb-Daimler-str. 46 67663 Kaiserslautern, Germany}
\address{$^2$present address: LNE-SYRTE, Observatoire de Paris, CNRS, UPMC, 61 avenue de l’Observatoire, 75014 Paris, France}
\address{$^3$present address: Laboratoire Kastler Brossel, ENS, UPMC-Paris 6, CNRS, 24 rue Lhomond, 75005 Paris, France}
\ead{tmanthey@physik.uni-kl.de}

%\date{\today}

\begin{abstract}
We report on the realization of high resolution electron microscopy of Rydberg-excited ultracold atomic samples. The implementation of an ultraviolet laser system allows us to excite the atom, with a single-photon transition, to Rydberg states. By using the electron microscopy technique during the Rydberg excitation of the atoms, we observe a giant enhancement in the production of ions. This is due to $l$-changing collisions, which broaden the Rydberg level and therefore increase the excitation rate of Rydberg atoms. Our results pave the way for the high resolution spatial detection of Rydberg atoms in an atomic sample.
\end{abstract}
%\maketitle
\input{introduction}
\input{exp_setup}

\input{UV_setup}
\input{SEM_UV}

\section*{References}
\bibliography{tech_ref}

\end{document}

%% file: introduction.tex
\section*{Introduction}

Because of their unique properties Rydberg atoms are an excellent tool for the investigation of few and many-body quantum systems. The large dipole moment ensures that locally only one excitation is possible, and thus a blockade arises \cite{Schwarzkopf2011,Schauss2012}. This leads to an excitation which is distributed over all atoms within the blockade volume, forming a collectively excited state \cite{Urban2009, Gaetan2009,Dudin2012}. Rydberg ensembles also strongly interact with light fields, which can be used to entangle atoms with photons for quantum communication purposes \cite{Li2013, Maxwell2013}. It has also been shown that the blockade effect can be used to implement quantum gates between neutral atoms \cite{Isenhower2010}. Another intriguing application is the dressing of ground state atoms with highly excited Rydberg states, giving rise to a tunable interaction potential between the dressed ground state atoms \cite{Honer2011, Johnson2010, Pupillo2010}. 

Direct Rydberg atom imaging has been demonstrated so far in optical lattices \cite{Schauss2012}, and in ion emission microscopy techniques \cite{Schwarzkopf2011}. Rydberg atom imaging capabilities give direct acces to spatial correlations and are the key component to study new quantum phases \cite{Cinti2010}, tailored spin systems \cite{Lee2011} and energy transfer mechanisms \cite{Wuester2011}.

Scanning electron microscopy has proven to be a powerful tool to manipulate and image ultracold atoms in their ground state \cite{Gericke2008, Wuertz2009, Gericke2007, Wuertz2006}. Due to the huge electric dipole moment Rydberg atoms are much more sensitive to electric fields and feature huge cross sections for electron scattering. Combining scanning electron microscopy with Rydberg excited atomic samples therefore offers new possibilities to prepare, image, and probe theses systems.

Our experimental apparatus combines a setup for the all-optical production of ultracold atomic samples with a scanning electron microscope for high resolution in-situ imaging and a high power ultra violet (UV) laser setup for one-photon Rydberg excitations. We first describe the apparatus and the experimental sequence that allows us to acquire scanning electron microscope images of Bose-Einstein condensates. Thereafter we describe the UV laser setup which allows the single-photon excitation of Rydberg atoms. In the last section we demonstrate the successful application of the electron microscopy technique on a BEC whose atoms are continuously excited to the $38P_{3/2}$ Rydberg state. The increase of the signal with respect to the non-excited case is far larger than the expected $n^2$ factor \cite{Vrinceanu2005}, making such technique promising  for the high resolution spatial detection of Rydberg atoms in atomic samples. With the help of a simple model we finally show that such giant enhancement is due to $l$-changing collisions induced by the electron beam itself.

%% file: exp_setup.tex
\section*{Experimental Setup}
\begin{figure}[btp]
\includegraphics[width=0.60\textwidth]{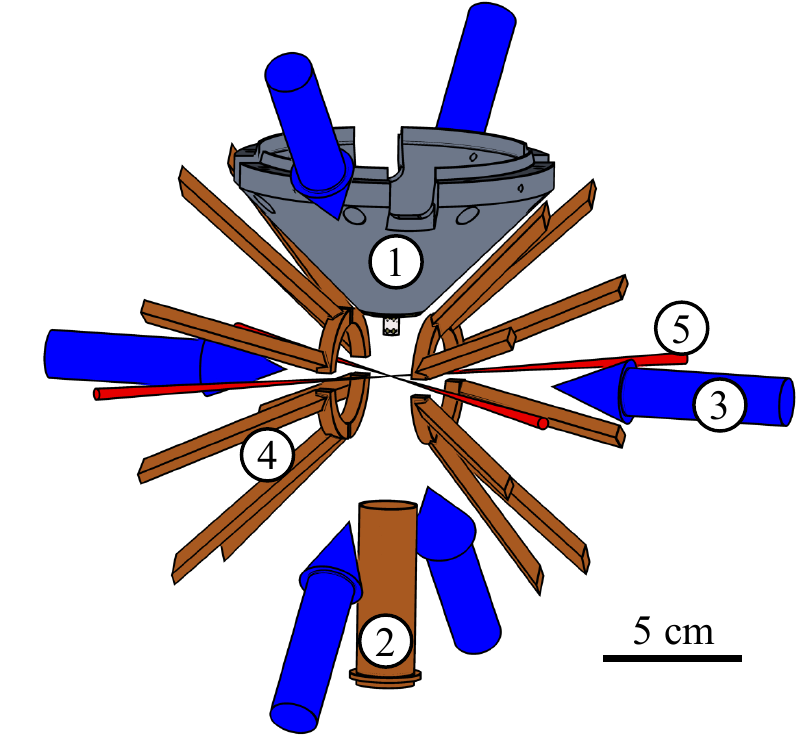}
\caption{Schematics of the interior of the science chamber. The electron column (1) is mounted on the top of the chamber. At the bottom of the science chamber the ion optics (2) is installed (only the first electrode is shown). The blue arrows show the six MOT beams (3). Eight electrodes are positioned near the center, forming two rings in Helmholtz configuration (4). The two laser beams (5) for the crossed dipole trap are crossing 11mm below the tip of the electron column, in the middle of the chamber.}
\label{fig:schematics}
\end{figure}

Our experimental sequence starts pre-cooling $^{87}$Rb atoms in a two dimensional magneto optical trap (2D MOT), orientated with an angle of 45$^\circ$ above the horizontal plane. By means of a resonant laser beam along the axis of the 2D MOT we transfer the atoms ($\unit[1 \times 10^9]{atoms/s}$) into the center of the science chamber, where they are collected and further cooled in a 3D MOT ($\unit[2.5 \times 10^9]{atoms}$ after $3s$). The laser sources for both traps are grating stabilized diode lasers locked on the D$_2$-transition line of Rubidium. The magnetic field for the 3D MOT is provided by eight electrodes placed directly inside the vacuum chamber. Each electrode has a section of $\unit[5 \times 5]{mm^2}$ and is made of oxygen-free copper. As shown in fig. \ref{fig:schematics} the electrodes are arranged in such a way that they form two effective coils in almost Helmholtz geometry with an inner radius of $\unit[15]{mm}$ and a distance of $\unit[25]{mm}$. The electrodes and the corresponding vacuum feed-throughs can hold up to $\pm \unit[1]{kV}$ and $\unit[200] {A}$. By setting the potential of the electrodes we can produce arbitrary electric fields ($\unit[500]{V/cm}$) or electric field gradients ($\unit[200]{V/cm^2}$) in the center of the chamber, which are of fundamental importance for the manipulation of Rydberg atoms. Alternatively magnetic fields ($\unit[70]{G}$) or magnetic field gradients ($\unit[20]{G/cm}$) can be produced by a current up to $\unit[200]{A}$ in the electrodes in Helmholtz or anti-Helmholtz configuration. During the 3D MOT the electrodes are driven in anti-helmholtz configuration with a current of $\unit[100]{A}$ which results in a magnetic field gradient of $\unit[10]{G/cm}$.

The subsequent evaporative cooling stage, necessary to reach the ultra-cold regime, is performed in a red detuned crossed dipole trap at $\unit[1064]{nm}$. The trapping light is provided by a fiber amplifier (Nufern, NuAmp fiberlaser) pumped with a low noise diode laser (INNOlight, Mephisto S). The light at the output of the amplifier is divided in three beams, a strong beam with a power of $\unit[15]{W}$ on the atoms and two weaker beams with $\unit[1.4]{W}$. The power of the strong beam is controlled with a free space AOM (Gooch \& Housego, R23080-3-1.06-LTD) and is focused on the atoms with a waist of $\unit[25]{\mu m}$ in the middle of the science chamber. To control the power of the weak beams we use two fiber-coupled AOMs (Gooch \& Housego, FiberQ). The two beams at the output of the fibers are focused to a waist of $\unit[80]{\mu m}$ in the center of the science chamber. One of them perpendicularly crosses the strong beam thus realizing a crossed dipole trap scheme. The other one is aligned in order to propagate collinearly with respect to the strong beam.

The transfer of the atoms from the 3D MOT to the crossed dipole trap is done in a dark MOT stage of $\unit[100]{ms}$, where we strongly suppress the repumping light to $1/70$ of its maximal power and widely detune the cooling light to $\unit[-195]{MHz}$. The strong and the weak orthogonal trapping beams are set to full power. At the end of the dark MOT stage we end up with $\approx \unit[1.5\times 10^7]{}$ atoms in the crossed dipole trap at a temperature of $\unit[250]{\mu K}$ in the $\vert F=1\rangle$ hyperfine manifold. After plain evaporation for $\unit[50]{ms}$ we start the forced evaporative cooling by exponentially lowering the power of the stronger beam while we hold the weak beam constant. This allows us to keep the collisional rate sufficiently high to ramp down the strong power beam with a time constant of $\tau_h=\unit[0.5]{s}$. After $\unit[1.9]{s}$ we also ramp down the low power beam with a time constant of $\tau_l=\unit[1.2]{s}$. Finally we end up with $\unit[2.5\times 10^5]{}$ atoms in a spinorial BEC after an evaporation time of $\unit[4]{s}$ . The final trapping frequencies are $\omega_x / \omega_y / \omega_z=  2\pi \times \unit[270/80/285]{Hz}$. 

In case a BEC in a defined polarized state is needed, we can apply a small magnetic field gradient along the vertical axis during the evaporation. This compensates the effect of gravity for the $|m_F=+1\rangle$ state while it enhances the effect of the gravity for the $|m_F=-1\rangle$ state. During the evaporation the atoms in the $|m_F=+1\rangle$ state always feel a deeper trap allowing us to selectively remove the atoms in the other states thus producing a purely polarized BEC ($\unit[1.0\times 10^5]{}$ atoms).
 
In order to get an isotropic trapping potential, we switch off the strong beam and switch on the third beam. This is done by sigmoidally ramping it up to the same power as the crossed beam in a time of $\unit[100]{ms}$, while the strong beam is completely ramped down in the last $\unit[100]{ms}$ of the evaporation stage. With this technique we end up with $\unit[2.5\cdot 10^5]{}$ atoms in a trapping potential with the frequencies $\omega_x / \omega_y / \omega_z= 2\pi \times \unit[80/80/94]{MHz}$.

\section{The scanning electron microscope}

\begin{figure}[tb]
\includegraphics[width=\textwidth]{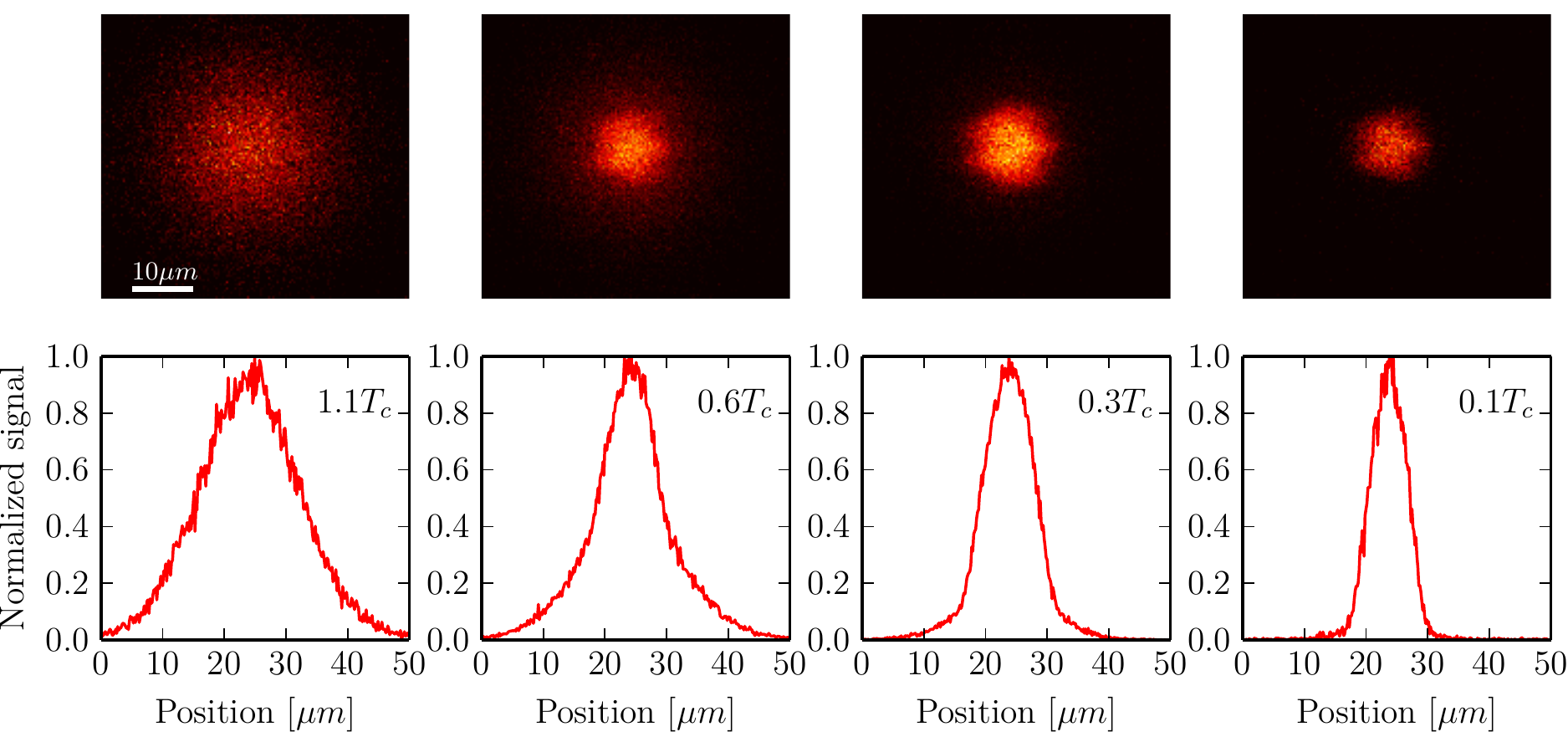}
\caption{In situ scanning electron microscopy images (summed over 50 runs) of an atomic sample across the BEC phase transition ($T_c\approx \unit[200]{nK}$) in a crossed dipole trap. The bottom row show the corresponding integrated signal along the vertical axis of the image.}
\label{fig:BEC1}
\end{figure}

On the upper part of our vacuum chamber we have installed a UHV-compatible scanning electron microscope (Delong instruments, ECLEX III). It consists of an emitter, placed on the upper end, which provides an electron beam (EB) with an energy of $\unit[6]{keV}$, and of two magnetic lenses. The current and the diameter of the EB can be set by different apertures, which can be inserted into the beam inside the column, and by the current of the first magnetic lens (gun lens). The second magnetic lens (objective lens) is used to focus the beam onto the sample. For beam shaping and guiding, three electric octopoles and one magnetic deflector are installed. The first electric octopole (located at the gun lens) and the magnetic deflector (located at the objective lens) are used to guide the beam through the optics and to correct for aberrations. The other two electric octopoles, installed at the very end of the column, are instead used to deflect the EB. The first, faster one ($\unit[200]{MHz}$), is used to move the beam along a scanning pattern, while the second, slower, is used to center the scan pattern on the atoms. Since it is possible to change the voltage of the octopoles in both directions independently, arbitrary scan patterns can be realized.

The cone of the electron column is placed inside the vacuum chamber and has a distance of $\unit[11]{mm}$ to the geometric center of the experiment (see fig \ref{fig:schematics}). The EB at the output of the electron microscope has a transverse 2D gaussian envelope. It is focused to the middle of the chamber, where it has a width (FWHM) of $\unit[170]{nm}$ with a beam current of $\unit[20]{nA}$ and a depth of focus of $\unit[35]{\mu m}$. The beam is finally stopped in a Faraday cup placed at the bottom of the vacuum chamber.

A high resolution electron-microscopy image of the trapped atomic cloud is obtained exploiting the ionization of the atoms produced by the electron-atom collisions \cite{Gericke2008}. Each time an electron collides with an atom there is $\simeq$40\% of probability to ionize it \cite{Wuertz2006}. The ions that are produced by electron-impact are guided by dedicated ion-optics to an ion detector (dynode multiplier, ETP 14553) placed at the bottom of the vacuum chamber. Synchronizing the arrival times of the ions with the scanning pattern of the EB we can reconstruct the \emph{in-situ} profile of the atomic density distribution \cite{Gericke2008}. In fig. \ref{fig:BEC1} we show scanning electron microscopy images of ultracold samples across the BEC transition together with the integrated density profiles. Due to the high resolution of $\unit[170]{nm}$ it is possible to observe the sample of atoms directly in the trap, which allows us to see the thermal fraction of the atomic sample at temperatures, where the time of flight imaging shows a pure BEC.

\begin{figure}
	\centering
		\includegraphics{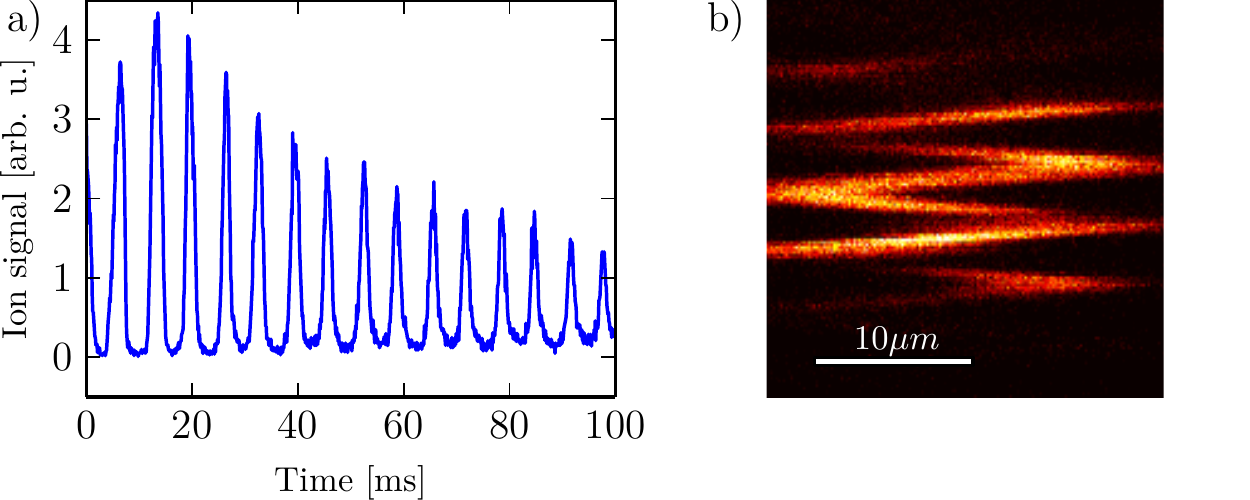}
	\caption{\textbf{a)} Ion signal during the oscillation of a BEC when the EB is aligned to the center of the trap. The BEC was displaced from the center of the trap with an additional trapping beam. After switching off this beam the BEC begins to oscillate with the trap frequency. From the clearly visible oscillation the trapping frequency is calculated via Fourier transformation. \textbf{b)} In situ scanning electron microscopy image of an moving BEC in an optical dipole trap. The scan starts at the bottom and then moves up in \unit[100]{ms}, while the atomic sample is oscillating.}
	\label{fig:trapfrequencies}
\end{figure}

\section*{Time resolved scanning electron microscopy}
Scanning electron microscopy has a sequential image formation process. This can be used to perform \textit{in vivo} studies of the time evolution of a quantum gas. For instance, when the EB is pointed at a fixed position in an atomic sample, the time evolution of the local atomic density can be examined. This can be directly used to measure the trapping frequencies very efficiently. By initially displacing one of the three trapping beams against the center of the optical dipole trap, we induce an oscillation of the BEC after abruptly switching off displaced beam. This oscillation can be directly monitored by the EB, which is aligned to the center of the trap and creates ions whenever the oscillating BEC passes underneath it.

In fig. \ref{fig:trapfrequencies}a we show the corresponding ion signal. The oscillation is clearly visible and the Fourier transform of the signal immediately gives the trapping frequency. In comparison to other techniques only a few experimental cycles are needed to determine the trapping frequencies. Alternatively, one can continuously perform a rectangular scan of the sample, thus visualizing the oscillations in space (fig. \ref{fig:trapfrequencies}b).

%% file: UV_setup.tex
\section*{Spectroscopy of the single photon transition}
To excite rubidium ground state atoms into Rydberg states, usually a two or three photon transition is used \cite{Bendkowsky2009, Maxwell2013, Weber2012}. In our setup we employ a one-photon transition, which has the advantage that no intermediate state is populated. Furthermore the excitation in nonisotropic P-states can be used to realize direction dependent interactions \cite{Reinhard2007}. In the case of Rubidium the transition wavelength to Rydberg states is around $\unit[297]{nm}$. The implementation of a powerful and tunable laser source at such wavelength is technically challenging. Our approach consists in frequency-doubling the light at $\unit[594]{nm}$ produced by a dye laser (Sirah, Matisse DS). Such laser, when pumped with $\unit[15]{W}$ of $\unit[532]{nm}$ light (Spectra Physics, Millennia), is able to provide up to $\unit[2.5]{W}$ laser power. The output laser beam is coupled to a photonic crystal fiber and sent to a bow-tie cavity kept under vacuum. The non linear element that we employ for the frequency-doubling is a cesium-lithium-borate (CLBO) crystal, chosen for its high damage threshold. The measured finesse of the cavity is 370 and the conversion efficiency is 50\%, allowing us to obtain up to $\unit[700]{mW}$ of UV light with a relative RMS power fluctuation of $\unit[4]{\%}$. The stabilization of the cavity length is realized by using a Pound-Drever-Hall scheme. After the cavity the UV laser is power stabilized by an AOM (AA optoelectronic, MQ200-A 1,5-266.300) and guided to the chamber, where it is focused on the atoms with a typical waist of $\unit[100]{\mu m}$.

An important feature of the UV laser setup is the wide tunability. We are able to excite every Rydberg level between $n=30$ and the ionization threshold. This is provided by the extremely large tuning range ensured by the dye laser. Its frequency is stabilized using an external cavity whose length is set by using an auxiliary diode laser at 780 nm. The frequency of this auxiliary laser is offset locked with respect to the cooling line of Rb. This allows us to tune the frequency of the UV laser in a range of up to $\unit[2]{GHz}$ without changing the locking point.

\begin{figure}
\includegraphics{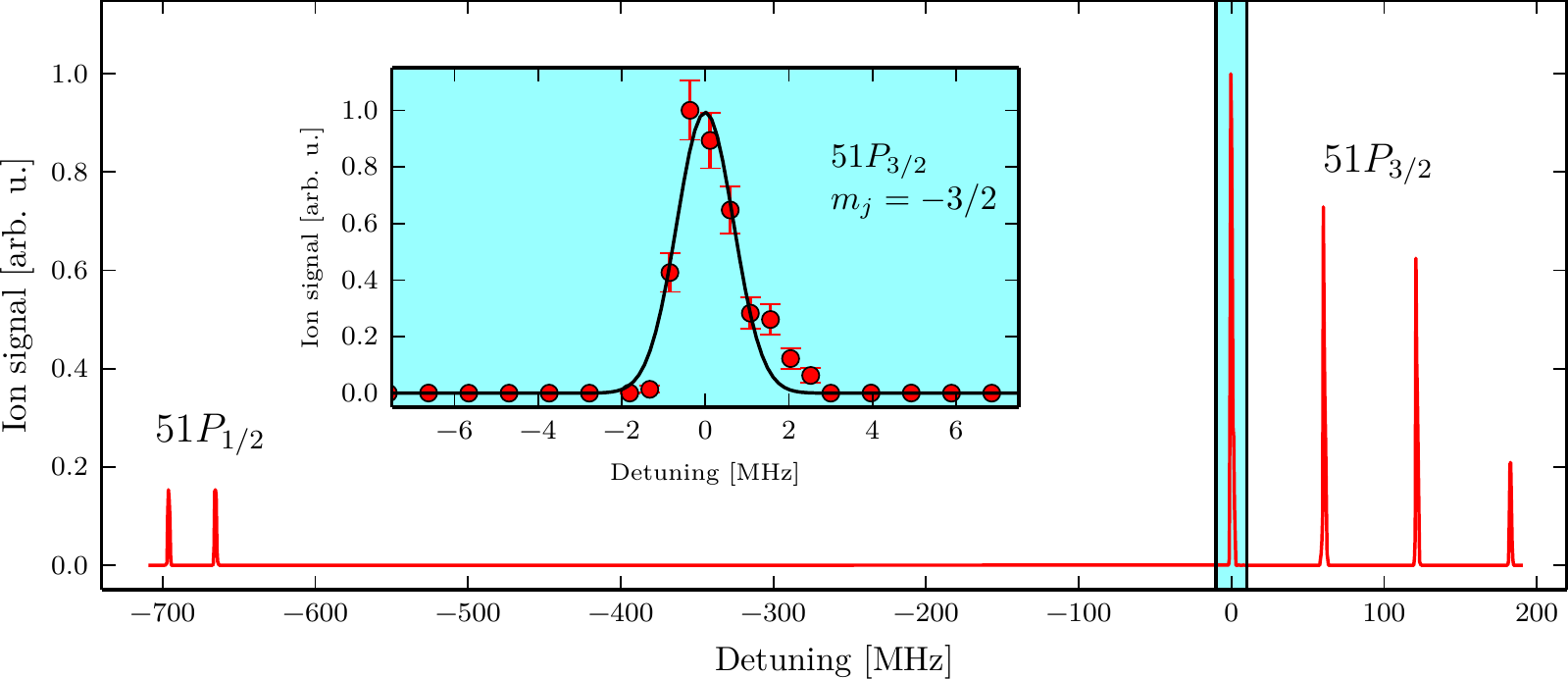}
\caption{Ion signal (red line) of the $5S \rightarrow 51P$ transition. The $m_j$ are splitted by a magnetic field of $\unit[34]{G}$. The inset shows the data (red dots) and a Gaussian fit (black line), which gives a linewidth of $\unit[0.7]{MHz}$ }
\label{fig:resonace}
\end{figure}

When atoms are excited to Rydberg states, several ionization channels are possible. In a dense ultra cold cloud they can evolve into ions via dipole-dipole interaction \cite{Li2006} or photo ionization due to the trapping laser \cite{Saffman2005} or black body radiation \cite{Beterov2009}. In the case of photo ionization the amount of ions is proportional to the excitation probability to the Rydberg state and is a direct measure of the excitation rate. In fig. \ref{fig:resonace} such an ion signal is shown as a function of the excitation laser detuning from the $5S_{1/2} (F=1) \rightarrow 51P$ transition. We have applied a magnetic field of $\unit[34]{G}$ to separate the individual $m_j$ states. The resonance was measured in a dilute gas with a peak density of $n= \unit[4.2\times 10^{11}]{cm^{-3}}$ and a temperature of $\unit[1.5]{\mu K}$. The spectrum illustrates the large mode-hop free tuning range of the UV laser setup and the small laser linewidth of less than 700\,kHz.

%% file: SEM_UV.tex
\section*{Electron-microscopy of a Rydberg excited BEC}

\begin{figure}[b]
\subfigure{\includegraphics[width=0.49\textwidth]{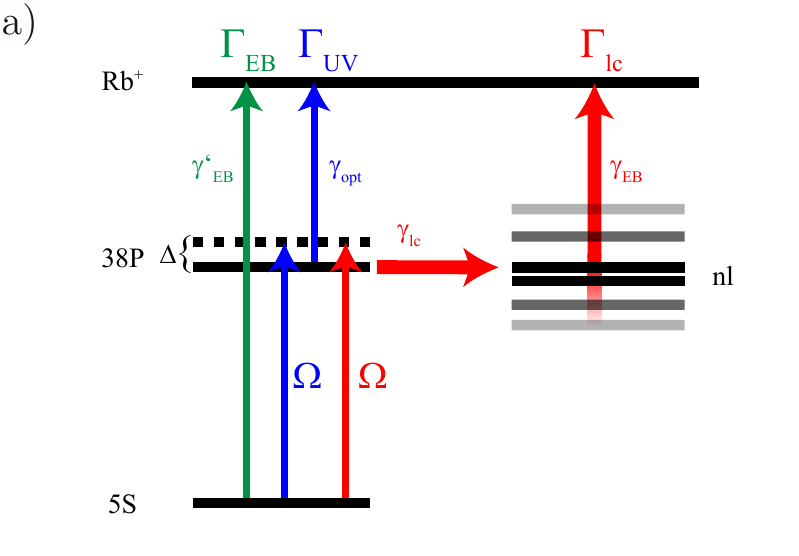}}\hfill
\subfigure{\includegraphics[width=0.49\textwidth]{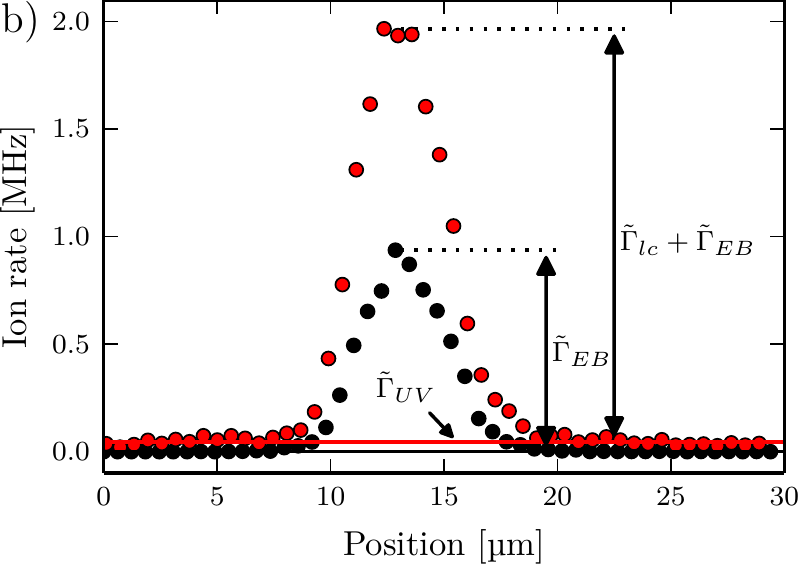}}
\caption{\textbf{a}) Schematics of the three ionization channels for a single atom. The first process (effective ion rate $\Gamma_{EB}$, green) is direct electron impact ionization of ground state atoms with ionization rate $\gamma'_{EB}$. The second ionization channel (effective ion rate $\Gamma_{UV}$, blue) is a two step process, consisting of an optical excitation with Rabi frequency $\Omega$ followed by an ionization process with rate $\gamma_{opt}$. The third ionization channel (effective ion rate $\Gamma_{lc}$, red) is a three step process, which consists of an optical excitation, an $l$-changing collision (rate $\gamma_{lc}$) and electron impact ionization of the Rydberg states (rate $\gamma_{EB}$). \textbf{b)} Ion rates during an electron beam line scan (waist $w=\unit[170(20)]{nm}$, current $I=\unit[25(1)]{nA}$) over a BEC with $N_{tot}=150$k atoms. The red (black) dots show the signal with (without) the simultaneous irradiation of the UV laser ($I=7\,$mW, $w=700\,\mu$m, $\Delta=2\pi\times10\,$MHz). The given rates correspond to the processes described in a) - the tilde indicates that they denote the integrated signal coming from all atoms which contribute to the signal. The ion rates are calculated with time bins of $\unit[10]{\mu s}$ corresponding to a position bin of $\unit[300]{nm}$.}
\label{fig:enhancement}
\end{figure}

In this section we investigate the effect of the single-photon Rydberg excitation on the performances of the electron microscopy. To do so we perform continuous line scans over a BEC with the electron beam alone and compare this with a line scan where we additionally shine the UV laser on the atomic sample. The UV beam is blue-detuned with respect to the $5S_{1/2}\rightarrow 38P_{3/2}$ transition. This is necessary to ensure a long enough lifetime and to avoid the excitation of molecular states on the red-detuned side \cite{Bendkowsky2009, Butscher2011}. In this configuration three different single-atom ionization channels are possible and they are illustrated in fig. \ref{fig:enhancement}a: the direct electron impact ionization of ground state atoms, the ionization caused by the UV laser alone and the ionization stemming from the combined action of the Rydberg excitation and of the EB. We define the corresponding single-atom ion rates as $\Gamma_{EB}$, $\Gamma_{UV}$ and $\Gamma_{lc}$ respectively. The effect of the single processes can be identified analyzing the line-scans reported in fig. \ref{fig:enhancement}b, where the UV light detuning is $\Delta=2\pi\times10$ MHz. The tilde on the symbols indicates that the measured ion rates are resulting from all the atoms present in the respective interaction volumes. The action of the UV light leads to a constant offset in the line scan of the Rydberg excited sample ($N_{tot}=150k$ atoms), as shown in fig. \ref{fig:enhancement}b, and the single-atom ion rate is calculated to be $\Gamma_{UV}=\tilde\Gamma_{UV}/N_{tot}=0.3$ Hz. When the EB hits the cloud and the UV beam illuminates the sample, an additional ion rate $\tilde\Gamma_{lc}$ is measured. In order to quantify and characterize this effect, in the following we analyze the ion rates measured when the EB illuminates the center of the cloud. The atoms ionized directly or indirectly by the EB are only those present in the small volume defined by the intersection of the atomic distribution and the EB itself. We calculate the number of atoms involved by integrating the atomic density over the volume of the EB, which we approximate by a cylinder with a radius of 170nm. In the center of the atomic distribution we find $N_{EB}\simeq360$ atoms. In such position the single-atom ion rate is $\Gamma_{lc}=\tilde\Gamma_{lc}/N_{EB}=2.7$ kHz while $\Gamma_{EB}=\tilde\Gamma_{EB}/N_{EB}=2.6$ kHz. Thus we observe that the ion rate for a single atom under off-resonant illumination with a UV laser can be enhanced by 4 orders of magnitude if an electron beam is shone simultaneously on the atoms ($\Gamma_{lc}/\Gamma_{UV}=0.9\times 10^4$). In addition, with respect to the electron impact ionization of the ground-state we observe an enhancement of a factor of $(\Gamma_{lc}+\Gamma_{EB})/\Gamma_{EB}=2$.

We now show that the observed giant enhancement can be explained by $l$-changing collisions in nearby Rydberg states. In this process the incoming electrons inelastically collide with Rydberg excited atoms and change the $n$ and $l$ quantum number. The scattering cross section for this event is given by \cite{Omivar1972}
\begin{equation}
\sigma_{n,l}=\pi a_0^2\frac{MZ^2R}{m_eE}\left[A(n,l)\ln\left( \frac{m_e}{M}\frac{1}{Z}\frac{E}{R}\right)+B(n,l)\right], 
\end{equation}
where $a_0$ is the Bohr radius, $m_e$ the electron mass, $M$ the mass of the atoms, $Z$ the charge of the nucleus, $E$ the collision energy and $R$ the ionization energy. The first coefficient $A(nl)$ is defined as \cite{Omivar1972}
\begin{equation}
A(nl) = \frac{1}{3} n^2\left\{n^2-1+3\left(l^2+l+4\right\}\right),
\end{equation}
which only depends on the quantum numbers $n$ and $l$. The second constant $B(nl)$ can be estimated, at least for high $n$ states, as an average over all the $l$ states: $B(n)/A(n) = a \ln(n) +b $ with $a\approx 1.84$ and $b\approx 2.15$.

With an EB energy of $\unit[6]{keV}$ we can calculate the $l$-changing cross section to be $\sigma_l (n=38,l=1)= \unit[8.7\times 10^{-13}]{cm^2}$. The resulting scattering rate into nearby Rydberg states is then given by
\begin{equation}
\gamma_{lc} = \sigma_l j /e ,
\end{equation}
where $e$ is the electron charge and $j=\unit[27.5]{A/cm^2}$ the current density of the EB. We obtain a scattering rate of $\gamma_{lc} = \unit[150]{MHz}$. This is much larger than every other decay or transition rate of the atom and therefore dominates the dynamics of the excitation process. For nearby states the $l$-changing scattering cross section is of the same order of magnitude so that excited atoms are frequently scattered between highly excited states due to the electron bombardment. Note that this inelastic scattering cross section is two orders of magnitude larger than the electron impact ionization cross section for the Rydberg state, which can be estimated to be around $\sigma_{ion} = \unit[4 \times 10^{-15}]{cm^2}$ \cite{Wuertz2010, Vrinceanu2005}. Once an atom is excited into a Rydberg state, it is several hundred times scattered in neighboring Rydberg states, before it is eventually ionized. As the electron impact ionization $\gamma_{EB}$ is by itself larger than the internal decay rates of the Rydberg state into low-lying states, we conclude that every atom which is excited to a Rydberg state is also ionized. 

For the calculation of the ion rate per atom in dependence on the detuning we use the stationary solution of the optical Bloch equation

\begin{equation}
\Gamma_{lc}(\Delta) = \frac{\gamma_{lc}}{2} \cdot \frac{\Omega^2/2}{\gamma_{lc}^2/4+\Omega^2/2+\Delta^2},
\label{eq:OBE_EB}
\end{equation}

with a Rabi frequency of $\Omega=\unit[2\pi \times 90]{kHz}$. We set the linewidth equal to $\gamma_{lc}$ as this is the dominating line broadening mechanism. The ion rate per atom for illumination with the UV beam alone is also given by the optical Bloch equation

\begin{equation}
\Gamma_{UV}(\Delta) = \frac{\gamma_{opt}}{2} \cdot \frac{\Omega^2/2}{\gamma_{w}^2/4+\Omega^2/2+\Delta^2},
\label{eq:OBE_UV}
\end{equation}

with a line width of $\gamma_w \ll \Delta$ (fig.\,\ref{fig:resonace}). The value of $\gamma_{opt}= \unit[50(10)]{kHz}$ is retrieved from a fit to the experimental data and summarizes all processes which transform the Rydberg excitation into ions. 

\begin{figure}[tb]
\includegraphics{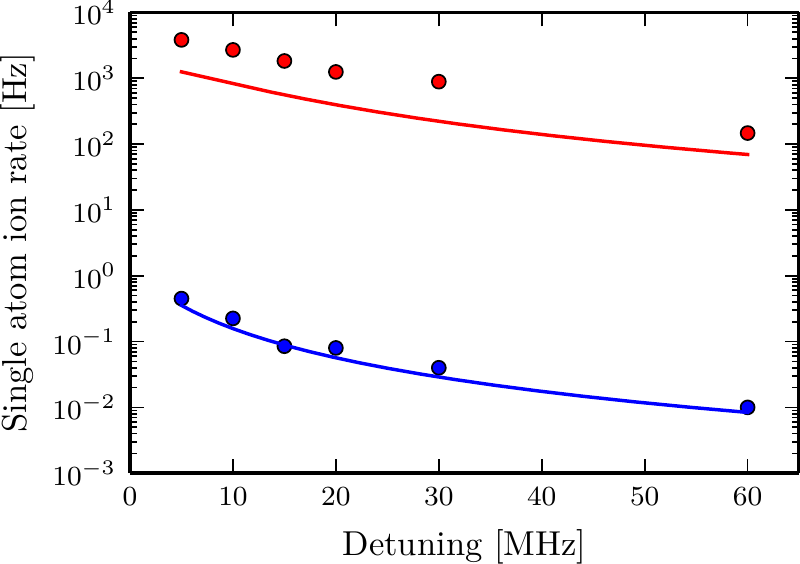}\hspace{2pc}
\begin{minipage}[b]{14pc}\caption{The ion rate per atom $\Gamma_{lc}$ (red dots) and $\Gamma_{UV}$ (blue dots) for different detunings in a logarithmic representation. The red (blue) line is the theoretical prediction of Eq.\,\ref{eq:OBE_EB} (Eq.\,\ref{eq:OBE_UV})}
\label{fig:logplot}
\end{minipage}
\end{figure}

In fig. \ref{fig:logplot} we summarize our findings. Indeed, the difference of four orders of magnitude between the two ionization processes with and without electron beam is well reflected by the theoretical treatment. It shows that electron impact scattering can be a powerful tool to enhance the excitation into Rydberg states. This can be used to create dissipative hot spots in Rydberg gases. In addition, it can be also employed to enhance the detection efficiency of scanning electron microscopy of ultracold quantum gases.

\section*{Outlook}
We have reported on the first experimental realization of electron microscopy of a Rydberg-excited atomic cloud. The experimental techniques employed allow us to create ultracold quantum gases under the emission tip of an electron microscope with cycle times of down to \unit[4]{s}, making our approach suitable for high statistical measurements. The electron microscope is a powerful tool for the manipulation and detection of atomic samples. Through the depletion of neighboring sites in an optical lattice it is possible to prepare very small samples to investigate isolated superatoms \cite{Lukin2001,Honer2011}. It is also possible to make \textit{in situ} and \textit{in vivo} measurements of an ultra cold sample. Our setup is equipped with a high-power UV laser source which allows us to drive single-photon transitions to Rydberg states. We have demonstrated that performing electron microscopy on samples coupled to Rydberg states greatly enhances the detection probability with respect to the non-excited case. Such an enhancement is even more pronounced than what is expected from the simple increase of the electron impact cross section of Rydberg atoms. We demonstrate that the giant enhancement is explained by $l$-changing collisions which take place when the atoms are bombarded by electrons and UV photons. For a pulsed electron beam, an extension to directly imaging the distribution of Rydberg atoms in an atomic gas is an intriguing vision.

We thank the DFG for the financial support within the SFB/ TRR 49. V.G and G.B. were supported by Marie Curie Intra-European Fellowships. 

%% file: main.bbl
\providecommand{\newblock}{}
\begin{thebibliography}{10}
\expandafter\ifx\csname url\endcsname\relax
  \def\url#1{{\tt #1}}\fi
\expandafter\ifx\csname urlprefix\endcsname\relax\def\urlprefix{URL }\fi
\providecommand{\eprint}[2][]{\url{#2}}
% Bibliography created with iopart-num v2.1
% /biblio/bibtex/contrib/iopart-num

\bibitem{Schwarzkopf2011}
Schwarzkopf A, Sapiro R~E and Raithel G 2011 {\em Phys. Rev. Lett.\/} {\bf 107}
  103001

\bibitem{Schauss2012}
Schauß P, Cheneau M, Endres M, Fukuhara T, Hild S, Omran A, Pohl T, Gross C,
  Kuhr S and Bloch I 2012 {\em Nature\/} {\bf 491} 87--91

\bibitem{Urban2009}
Urban E, Johnson T~A, Henage T, Isenhower L, Yavuz D~D, Walker T~G and Saffman
  M 2009 {\em Nature Physics\/} {\bf 5} 110--114

\bibitem{Gaetan2009}
Gaëtan A, Miroshnychenko Y, Wilk T, Chotia A, Viteau M, Comparat D, Pillet P,
  Browaeys A and Grangier P 2009 {\em Nature Physics\/} {\bf 2} 115 -- 118

\bibitem{Dudin2012}
Dudin Y~O, Li L, Bariani F and Kuzmich A 2012 {\em Nature Physics\/} {\bf 8}
  790--794

\bibitem{Li2013}
Li L, Dudin Y~O and Kuzmich A 2013 {\em Nature\/} {\bf 498} 466--469

\bibitem{Maxwell2013}
Maxwell D, Szwer D, Paredes-Barato D, Busche H, Pritchard J, Gauguet A,
  Weatherill K, Jones M and Adams C 2013 {\em Phys. Rev. Lett.\/} {\bf 110}
  103001

\bibitem{Isenhower2010}
Isenhower L, Urban E, Zhang X~L, Gill A, Henage T, Johnson T, Walker T and
  Saffman M 2010 {\em Phys. Rev. Lett.\/} {\bf 104} 010503

\bibitem{Honer2011}
Honer J, Löw R, Weimer H, Pfau T and Büchler H~P 2011 {\em Phys. Rev. Lett\/}
  {\bf 107} 093601

\bibitem{Johnson2010}
Johnson J~E and Rolston S~L 2010 {\em Phys. Rev. A\/} {\bf 82} 033412

\bibitem{Pupillo2010}
Pupillo G, Micheli A, Boninsegni M, Lesanovsky I and Zoller P 2007 {\em Phys.
  Rev. A\/} {\bf 75} 032712

\bibitem{Cinti2010}
Cinti F, Jain P, Boninsegni M, Micheli A, Zoller P and Pupillo G 2010 {\em
  Phys. Rev. Lett\/} {\bf 105} 135301

\bibitem{Lee2011}
Lee T~E, Häffner H and Cross M~C 2011 {\em Phys. Rev. A\/} {\bf 84} 031402

\bibitem{Wuester2011}
Wüster S, Ates C, Eisfeld A and Rost J~M 2011 {\em New J. Phys\/} {\bf 13}
  073044

\bibitem{Gericke2008}
Gericke T, Würtz P, Reitz D, Langen T and Ott H 2008 {\em Nature Physics\/}
  {\bf 4} 949--953

\bibitem{Wuertz2009}
Würtz P, Langen T, Gericke T, Koglbauer A and Ott H 2009 {\em Phys. Rev.
  Lett\/} {\bf 103} 080404

\bibitem{Gericke2007}
Gericke T, W\"urtz P, Reitz D, Utfeld C and Ott H 2007 {\em Appl. Phys. B\/}
  {\bf 89} 447--451

\bibitem{Wuertz2006}
Würtz P, Gericke T, Langen T, Koglbauer A and Ott H 2006 {\em J. Phys: Conf.
  Ser.\/} {\bf 141} 012020

\bibitem{Vrinceanu2005}
Vrinceanu D 2005 {\em Phys. Rev. A\/} {\bf 72} 022722

\bibitem{Bendkowsky2009}
Bendkowsky V, Butscher B, Nipper J, Shaffer J, Löw R and Pfau T 2009 {\em
  Nature\/} {\bf 458} 1005--1008

\bibitem{Weber2012}
Weber T, Niederprüm T, Manthey T, Langer P, Guarrera V, Barontini G and Ott H
  2012 {\em Phys. Rev. A\/} {\bf 86} 020702

\bibitem{Reinhard2007}
Reinhard A, Liebisch T~C, Knuffman B and Raithel G 2007 {\em Phys. Rev. A\/}
  {\bf 75} 032712

\bibitem{Li2006}
Li W, Tanner P, Jamil Y and Gallagher T 2006 {\em Eur. Phys. J. D\/} {\bf 40}
  27--35

\bibitem{Saffman2005}
Saffman M and Walker T~G 2005 {\em Phys. Rev. A\/} {\bf 72} 022347

\bibitem{Beterov2009}
Beterov I~I, Tretyakov D~B, Ryabtsev I~I, Entin V~M, Ekers A and Bezuglov N~N
  2009 {\em New J. Phys\/} {\bf 11} 013052

\bibitem{Butscher2011}
Butscher B, Bendkowsky V, Nipper J, Balewski J~B, Kukota L, Löw R, Pfau T, Li
  W, Pohl T and Rost J~M 2011 {\em J. Phys. B: At. Mol. Opt. Phys.\/} {\bf 44}
  184004

\bibitem{Omivar1972}
Omivar K and Khateeb A 1973 {\em J. Phys. B\/} {\bf 6} 341--353

\bibitem{Wuertz2010}
Würtz P, Gericke T, Vogler A and Ott H 2010 {\em New J. Phys\/} {\bf 12}
  065033

\bibitem{Lukin2001}
Lukin M~D, Fleischhauer M, Cote R, Duan L~M, Jaksch D, Cirac J~I and Zoller P
  2001 {\em Phys. Rev. Lett\/} {\bf 87} 037901

\end{thebibliography}
